\documentclass[prb,
superscriptaddress,showpacs,amsmath,amssymb]{revtex4}

\begin{document}

\title{Hawking radiation: black hole vs de Sitter}

\author{G.E.~Volovik}
\affiliation{Landau Institute for Theoretical Physics, acad. Semyonov av., 1a, 142432,
Chernogolovka, Russia}

\date{\today}

\begin{abstract}
We discuss the difference between the thermodynamics of black holes and thermodynamics of the de Sitter expansion. Both systems experience the Hawking radiation, but its impact on  thermodynamics is different. As distinct from the  thermodynamics of black holes, which are finite compact objects, the de Sitter state is the infinite and homogeneous state. The presence of the cosmological horizon provides  two sides of the de Sitter thermodynamics: the local thermodynamics and the thermodynamics related to the cosmological horizon. We discuss the connection between these two sides considering the entropy of the Hubble volume in de Sitter spacetime -- the region inside the horizon. On one hand there is
 the Gibbons-Hawking entropy $S_{\rm GH}=A/4G$ associated with the cosmological horizon. On the other hand this entropy can be obtained by integrating the local entropy density over the Hubble volume. In (3+1) spacetime, these two entropies coincide. This provides physical meaning and a natural explanation to the Gibbons-Hawking entropy  -- it is the entropy in the volume $V_H$ bounded by the cosmological horizon.
Here we consider whether the Gibbons-Hawking conjecture remains valid for the de Sitter state in general $d+1$ spacetime. To do this, we use the local de Sitter thermodynamics, characterized by a local temperature $T_{\rm dS}=H/\pi$. This temperature is not related to the horizon: it is the temperature of local activation processes, such as the ionization of an atom in the de Sitter environment, which occur deep within the cosmological horizon. This local temperature is universal and does not depend on dimension $d$, it is twice the Gibbons-Hawking temperature $T_{\rm GH}=H/2\pi$. We found that the entropy of the Hubble volume is $S_H=(d-1)A/8G$, which modifies the Gibbons-Hawking entropy of horizon. The original form of the Gibbons-Hawking entropy is valid only for $d=3$. 
\end{abstract}
\pacs{
}

\maketitle

\tableofcontents
 
 \section{Introduction}

It is now accepted that Bekenstein-Hawking entropy of a black hole is associated to the area of the event horizon, $S_{\rm BH}=A/4G$, although there are still some doubts.\cite{Hooft2024} Later it was suggested  by Gibbons and Hawking that the area of cosmological horizon can be also interpreted as an entropy.\cite{GH1977}
A black hole is a compact object, the position of the horizon is well determined, and in the asymptotically flat spacetime the concept of its entropy is physically motivated. In contrast, the de Sitter universe is a homogeneous state, and the position of the cosmological horizon depends on the position of the observer. That is why the meaning of the horizon entropy requires explanation.\cite{Diakonov2025}

The proper approach is to consider this surface entropy as coming from the bulk contribution, see Ref.\cite{Volovik2024} and references therein.  In this approach, the Sitter state is considered as a medium which is characterized by the local temperature $T_{\rm dS}=H/\pi$. This temperature describes the local processes in the de Sitter environment, such as the activation process of ionization of an atom, \cite{Volovik2009,Maxfield2022}  the decay of a composite particle, \cite{Bros2008,Bros2010,Volovik2009} and decay of massive fields into pairs of inflatons.\cite{Maldacena2015}   For example, the rate of ionization of an atom is
 $w \propto \exp{(-\epsilon/T_{\rm dS})}$, where $\epsilon$ is the ionization potential and  $T_{\rm dS}=H/\pi$. The decay rate of a particle with mass $m_0$ into two particles  of mass $m_1$ for $2m_1 > m_0$ is also determined by this local temperature, $w \propto \exp{(-\frac{2m_1-m_0}{T_{\rm dS}})}$. This local temperature is twice the Gibbons-Hawking temperature,\cite{GH1977} $T_{\rm GH}=H/2\pi$.

The local temperature $T_{\rm dS}=H/\pi$ determines the local entropy density $s(T)$. Integrating of the entropy density over the Hubble volume $V_H$ one obtains the entropy of the Hubble volume, which coincides with the horizon entropy $S_{\rm GH}$ suggested by Gibbons and Hawking,
$S_H=s(T)V_H=A/4G=S_{\rm GH}$. This coincidence gives physical meaning and a natural explanation to the  Gibbons-Hawking entropy of the horizon: it can be associated with the entropy of the region inside the cosmological horizon.

In Ref. \cite{Diakonov2025} the relationship between these two entropies (called on-shell and off-shell entropies) was considered for a general $(d+1)$ spacetime. As in our approach, the on-shell entropy -- the entropy of the Hubble volume -- was computed as a volume integral in de Sitter space. It was found that in any $(d+1)$-dimensional spacetime, the holographic bulk-horizon correspondence remains valid: the bulk contribution coincides with the Gibbons-Hawking entropy $S_{\rm GH}=A/4G$. However, in the calculations, instead of the local temperature $T=H/\pi$, the Gibbons-Hawking temperature $T_{\rm GH}=H/2\pi$ was used, which causes the problem.
  We extended our approach for general dimension $d$, and found that the entropy of the Hubble volume and thus the entropy associated with the horizon is modfied: $S_H=\frac{d-1}{8}A/G$. The modification of Gibbons-Hawking entropy is also confirmed by consideration of the first law of de Sitter thermodynamics,\cite{Volovik2025F} which is also based on the local temperature $T=H/\pi$.  

We start with Section \ref{BHSec}, which demonstrates how the temperature of Hawking radiation and the Bekenstein-Hawking entropy of black holes  arise from the consideration of the correspondingly quantum tunneling and macroscopic quantum tunneling.

In Section \ref{dSthermo}, the local thermodynamics of de Sitter is discussed.
It is shown that it looks similar to the Landau two-fluid hydrodynamics, where the role of the superfluid component is played by the dark energy, while gravitational degrees of freedom form the thermal component. In particular, in Section \ref{SecondSound} it is shown that the graviton propagating  in the de Sitter spacetime is obtained directly from the Landau equation for the velocity of the second sound -- the propagating mode of entropy density.
In Section \ref{QCDSec} the two-fluid thermodynamics of de Sitter is applied to the estimation of the present value of the cosmological "constant" using the Gribov approach to the QCD confinement.

In Section \ref{ModEntropySec} the global thermodynamics of the Hubble volume is discussed.
The first law of this thermodynamics is in Section \ref{FirstLawSec}.
In Section \ref{contracting} we discuss the controversy between our approach and that in Ref. \cite{Diakonov2025}, which is attributed to the sign problem.\cite{Jacobson2023} 
We show that the negative sign does appear, but only for the contracting de Sitter state, which has negative $H<0$, negative $T<0$ and correspondingly the negative entropy of the Hubble volume.
In Section \ref{WaldSec} we considered the possible extension of the Wald entropy to satisfy the modified value $S_H=\frac{d-1}{8}A/G$ of the Gibbons-Hawking entropy. 
Section \ref{fluctuations} is devoted to the quantum and classical fluctuations of the cosmological horizon in general dimension $d$. In Section \ref{CombineSec} we consider the application of the Tsallis-Cirto composition law to the cosmological horizon.

 \section{Black hole: Hawking temperature and Bekenstein-Hawking entropy}
  \label{BHSec}
  
 \subsection{Hawking radiation by quantum tunneling}
\label{HawkingRadiation}

It is convenient to study the  Hawking radiation from black hole and in de Sitter using the metric in the Painlev\'e-Gullstrand (PG) form:\cite{Painleve,Gullstrand}
\begin{equation}
ds^2= - dt^2 +   (d{\bf r} - {\bf v}({\bf r})dt)^2\,.
\label{PG1}
\end{equation}
Here ${\bf v}({\bf r})$ is the shift vector, which corresponds to the superfluid velocity in the condensed matter analogues. For the black hole one has $v^2(r)=2GM/r$, where $M$ is the black hole mass and $G$ is the Newton constant. In the de Sitter state the shift velocity is ${\bf v}({\bf r})=H{\bf r}$, and the metric is:
\begin{equation}
ds^2= - dt^2 +   (dr - Hr dt)^2+r^2 d\Omega^2\,.
\label{PG}
\end{equation}
 
In both cases the particle spectrum is given by
\begin{equation}
E({\bf p},{\bf r})= \pm c p + {\bf p}\cdot {\bf v} ~,
\label{Spectrum}
\end{equation}
where the first term is the spectrum in the frame  comoving with the vacuum, while the last term plays the role of the Doppler frequency shift
Equation $E({\bf p},{\bf r})=E={\rm const}$ determines the radial trajectories of massless particles: 
\begin{equation}
p_r(r)=\frac{E}{v(r)\pm c} ~.
\label{Trajectories}
\end{equation}
The integration contour contains the path along the semicircle around the pole at $r=r_g$ where $|v(r)|=c$, which relates the tunneling probability to the imaginary part in the semiclassical action $w\propto \exp(-2{\bf Im}~S)$:
\begin{equation}
2{\bf Im}~S=2{\bf Im}\int dr ~p_r(r)=2E~ {\bf Im}\int\frac{dr}{v(r)\pm c} = \frac{2\pi E} {dv/dr|_{r=r_g}}~.
\label{Action}
\end{equation}
The latter is interpreted as the thermal  radiation from the horizon, which is described  by the Hawking temperature
\begin{equation}
T_{\rm H}= \frac{\hbar}{2\pi } \frac{dv} {dr}\Big |_{r=r_g}~.
\label{HawkingT}
\end{equation}

For the black hole with mass $M$, this tunneling approach gives
the following probability of emission of the particle with energy $\omega$ per unit time:\cite{ParikhWilczek2000,Volovik1999} 
\begin{equation}
w(\omega, M)\propto   \exp{\left(-8\pi GM\omega\right)}\equiv \exp{\left(-\frac{\omega}{T_{\rm H}}\right)} 
\,,
\label{tunneling}
\end{equation}
where $T_{\rm H}$ is the temperature of the Hawking radiation:
\begin{equation}
T_{\rm H}=\frac{1}{8\pi GM}
\,,
\label{HawkingT}
\end{equation}
and we use  the condition $T_{\rm H}\ll \omega \ll M$. 
Here we use the units $\hbar=c=1$, however in some cases these parameters are restored for greater clarity, while in some cases we also use $G=1$ to simplify equations.

For the de Sitter Universe with its $v(r)=Hr$, the corresponding temperature
is the Gibbons-Hawking temperature of the cosmological horizon:
\begin{equation}
T_{\rm GH}= \frac{H}{2\pi }~.
\label{HawkingTDS}
\end{equation}
But while the Hawking temperature $T_{\rm H}$ determines the black hole thermodynamics, the Gibbons-Hawking temperature $T_{\rm GH}$ does not determine the thermodynamics of the de Sitter state. This is what distinguishes finite systems from infinite ones. We consider the de Sitter thermodynamics in Section \ref{dSthermo}, but now concentrate on black hole thermodynamics.

\subsection{Back reaction and the Bekenstein-Hawking entropy}
\label{BackReactionSec}

 Parikh and Wilczek\cite{ParikhWilczek2000} obtained the correction to the Hawking radiation, which is caused by the back reaction --  the reduction of the black hole mass after emission:
 \begin{equation}
w(\omega, M-\omega)\propto \exp{\left(-8\pi G\omega\left(M-\frac{\omega}{2}\right)\right)}
\,.
\label{tunnelingMomega}
\end{equation}
This back reaction opened another path to the concept of black hole entropy -- the Bekenstein-Hawking entropy. The process of Hawking radiation, described by the Hawking temperature, can be viewed as a thermodynamic fluctuation. According to Landau and Lifshitz,\cite{Landau_Lifshitz} the rare fluctuations lead to the decrease of entropy, and thus
the rate of Hawking radiation in Eq.(\ref{tunnelingMomega}) can be described in terms of the decrease of the  black hole entropy after emission of a particle:
\begin{equation}
w(\omega, M-\omega)\propto \exp{\left[S_{\rm BH}(M-\omega)-S_{\rm BH}(M)\right]}
\,.
\label{EntropyDifference}
\end{equation}
This immediately gives the value $S_{\rm BH}(M)=4\pi GM^2$ for  the Bekenstein-Hawking entropy, which agrees with the thermodynamic law $1/T_{\rm H}=dS_{\rm BH}/dM$.

This shows that the quantum tunneling processes serve as a source of both the Hawking temperature and the Bekenstein-Hawking black hole entropy. The entropy consideration is also supported by consideration of macroscopic quantum tunneling discussed in Sec. \ref{TunnelingEmission}. It describes the process in which a large black hole emits smaller black holes.\cite{Volovik2022G}  These macroscopic processes occurring in the ensemble of black and white holes obey the peculiar non-extensive Tsallis-Cirto statistics discussed in Sec. \ref{NonextensiveSec}.

\subsection{Emission of black holes as macroscopic quantum tunneling}
\label{TunnelingEmission}

The emitted black hole can be thought of as a type of particle, and the emission process contains a similar element of back reaction.\cite{Volovik2022G} However, such "particle" emitted by the black hole has a non-zero entropy. As a result, unlike the emission of a point particle, the emission rate of a small black hole increases by the entropy of the emitted black hole.\cite{HawkingHorowitz1995} 

All this suggests that the general process of the splitting of black holes into several parts can be expressed in terms of the entropies of the black holes participating in this process. In particular, the rate at which a black hole splits into two smaller black holes in the process of macroscopic quantum tunneling obeys the following rule:
 \begin{eqnarray}
w(M\rightarrow M_1+M_2)\propto 
\nonumber
\\
\propto\exp{\left[S_{\rm BH}(M_1)+S_{\rm BH}(M_2)-S_{\rm BH}(M_1+M_2)\right]}
\,.
\label{HoleEmession2}
\end{eqnarray}
This equation allows us to obtain the entropy of the black hole as a function of its mass, 
$S_{\rm BH}(M)$. For that let us consider the emission of a small black hole with mass $m$ by a large black hole with mass $M \gg m$.
The rate of emission obeys Eq.(\ref{tunneling}) with $\omega=m$. On the other hand, it obeys Eq.(\ref{HoleEmession2})
with $M_1=M-m$ and $M_2=m\ll M$. Expanding this equation in small $m$ and comparing this equation with Eq.(\ref{tunneling}) one obtains the function $S_{\rm BH}(M)$:
\begin{equation}
\frac{dS_{\rm BH}}{dM}=8\pi GM \,\, \rightarrow \,\, S_{\rm BH}(M)=4\pi GM^2\,.
\label{EntropyExpansion}
\end{equation}

The macroscopic quantum tunneling approach is actually another way to derive the Bekenstein-Hawking entropy of a black hole. Application of the macroscopic quantum tunneling to the  cosmological decay of a false quantum vacuum can be found in Refs.\cite{HawkingMoss1982,Saito2025}.

\subsection{Non-extensive entropy of black hole}
\label{NonextensiveSec}

The black hole entropy is non-extensive, since $S_{\rm BH}(M_1 +M_2) > S_{\rm BH}(M_1)+ S_{\rm BH}(M_2)$. The source of the non-extensive entropy of black holes is a special type of configuration space of the black hole ensemble, which follows from the black hole transformations discussed in Sec. \ref{TunnelingEmission}.

The entropy of the conventional thermodynamic systems is extensive. The entropy there is proportional to the volume of the system, and thus the splitting of the system with volume $V$ in two parts with volumes $V_1+V_2=V$ does not change the total entropy of the system,  $S(V_1+V_2)=S(V_1) +S(V_2)$, or $S(A,B)=S(A) +S(B)$. 

The black hole entropy does not satisfy the additivity condition. Instead one has the following non-additive composition rule for the black hole entropies:
\begin{equation}
S_{\rm BH}(M_1 +M_2)= \left( \sqrt{S_{\rm BH}(M_1)} + \sqrt{S_{\rm BH}(M_2)}\right)^2 \,.
\label{TwoBlackHoles}
\end{equation}

It is precisely because of the quantum processes that the ensemble of black holes has a special type of configuration space, where the entropy is not extensive. This non-extensivity is caused by quantum fluctuations, which determine the rare processes of macroscopic quantum tunneling between the black hole states. This is the analog of intermittency in the chaotic systems.\cite{Robledo2022}
Such processes require the generalization of the statistics with the corresponding non-extensive entropy. The equation (\ref{TwoBlackHoles}) fully determines the type of the statistics: it is described by the Tsallis-Cirto entropy with 
$\delta=2$,\cite{Volovik2025} see Section \ref{TsallisSec}.

If a black hole is a mixed state and its entropy is a kind of the von Neumann entropy, then the quantum tunneling process of breaking the black hole into smaller black holes transforms the mixed state into a less mixed state - a state with lower entropy. 
This is consistent with Weinberg's view\cite{Weinberg2014} that it is the density matrix introduced by Landau and von Neumann that describes physical reality, not the wave function introduced by Schr\"odinger. The wave function approach is as convenient as the complex order parameter (the condensate wave function) introduced by Ginzburg and Landau to describe superfluidity  and superconductivity. But the real physical quantities are their bilinear combinations - the density matrix and the correlation function, which are invariant with respect to phase shift (for additional symmetries of the density matrix compared to the wave function, see Ref. \cite{Weinberg2014}).

\subsection{Composition rule for merging and splitting of black holes}
\label{TsallisSec}

The black hole entropy $S_{\rm BH}(M)=4\pi GM^2$ is non-extensive with the special type of composition.  An example, which we considered, was the process of the splitting of the black hole with mass $M$ into two smaller black holes with masses $M_1$ and $M_2$. For $M_1+M_2=M$, the entropy in Eq.(\ref{TwoBlackHoles}) obeys the following composition rule:
\begin{equation}
\sqrt{S_{\rm BH}(M=M_1 +M_2)}= \sqrt{S_{\rm BH}(M_1)} +\sqrt{S_{\rm BH}(M_2)}\,.
\label{BlackHoles}
\end{equation}
This composition suggests  the application of the non-extensive Tsallis-Cirto $\delta=2$ entropy:\cite{TsallisCirto2013}
\begin{equation}
S_{\delta =2}=\sum_i p_i \left(\ln\frac{1}{p_i} \right)^2\,,
\label{TCentropy}
\end{equation}
which gives for a system composed of two probabilistically independent subsystems $A$ and $B$, the following non-additive composition rule:
\begin{equation}
\sqrt{S_{\delta =2}(A+B)}=\sqrt{S_{\delta =2}(A)} + \sqrt{S_{\delta =2}(B)}\,.
\label{TCentropy2}
\end{equation}
The non-extensive entropy is a direct consequence of the non-extensivity of the gravitational interaction, i.e. the consequence of the long-range force of gravity, see e.g.\cite{Landsberg1980,Landsberg1984,Padmanabhan1990,Nakamichi2002,Odintsov2025}. This can be seen if one considers the black hole with mass $M$ as ensemble of $N$ black holes with masses $m_0$, which gravitationally interact with each other.\cite{VolovikPlanckons} The entropy of this ensemble is proportional not to the number $N$ of the objects, but to the number $N(N-1)/2$ of ways to select two interacting objects from this ensemble. As a result, the entropy of black hole is proportional to $N^2=M^2/m_0^2$, which agrees with Bekenstein-Hawking entropy, if $m_0$ is on the order of the Planck mass. That is why statistically a black hole can be considered as the statistical ensemble of Planck scale black holes -- planckons  or maximons.\cite{Markov1967}

\subsection{Negative entropy of white hole}
\label{WhiteHoleSec}

The black and white holes differ by the opposite directions of the shift vector,
${\bf v}({\bf r})= \pm {\hat{\bf r}}\sqrt{2MG/r}$. The minus and plus signs correspond to the motion towards the central singularity in the black hole and away from the singularity in the white hole.
The sign reversal suggests that black and white holes are the physical objects with the opposite signs of the entropy.  The negative value of the white hole entropy can be obtained using the process of macroscopic quantum tunneling from a black hole to a white hole of the same mass, which rate is $w \propto e^{-8\pi GM^2}$. According to Landau and Lifshitz\cite{Landau_Lifshitz} this process can be considered as rare fluctuation which rate is determined by the difference in entropies in these two states, $w \propto e^{S_{\rm WH}-S_{\rm BH}}$. This demonstrates that the entropy of the white hole is equal to the entropy of the black hole with a minus sign
\begin{equation}
S_{\rm WH}(M)=-4\pi GM^2=-S_{\rm BH}(M)\,.
\label{WHentropy}
\end{equation}
In the further process of the relaxation of the white hole towards the black hole, the entropy increases, first reaching the zero value at the Schwarzschild black hole with coordinate singularity at the horizon and finally reaching the maximum value when the black hole is formed.

It was shown\cite{Volovik2025TC} that the composition law (\ref{BlackHoles}) can be extended to include the white holes with negative entropy and also to the inner horizons of black holes, which also have negative entropy. For example, if the black hole splits into the black hole with mass $M_1$ and the white hole with mass $M_2$, the Eq.(\ref{BlackHoles}) becomes
\begin{equation}
\sqrt{S_{\rm BH}(M=M_1 +M_2)}= \sqrt{S_{\rm BH}(M_1)} +\sqrt{|S_{\rm WH}(M_2)|}\,.
\label{BlackWhiteHoles}
\end{equation}
For $M_1=0$ and $M_2=M$ this corresponds to the process of macro-tunneling from black to  white hole.

The negative entropy of the contracting de Sitter with $H<0$ will be discussed in Sec. \ref{contracting}.

\subsection{Entropy of rotating black hole}
\label{KerrSec}

The black hole entropy is the consequence of quantum/thermodynamic statistics of the ensembles of the black holes, which determines the rate of different processes of merging and splitting of black and white holes. These processes are governed by the non-extensive Tsallis-Cirto  $\delta=2$ statistics.\cite{TsallisCirto2013} 
Let us apply this statistics to the composition law for the inner and outer horizons of the Kerr black hole.\cite{Volovik2026} According to this composition law the total entropy of the Kerr black hole is expressed in terms of the entropies of the inner and outer horizons in the same way as for the RN black hole,\cite{Volovik2025TC} where the entropy of the inner horizon is negative (the entropy of states with multiple horizons see also in Ref. \cite{Azarnia2024}). For the Kerr black hole one has 
\begin{equation} 
\sqrt{S_{\rm Kerr}}= \sqrt{S_+} +\sqrt{|S_-|}\,,
\label{BHTsallis}
\end{equation}
where the entropy of outer horizon is
\begin{equation}
S_+ = 2\pi GM^2 \left(1+\sqrt{1-J^2/M^4G^2}   \right)\,,
\label{outer}
\end{equation}
while the entropy of the inner horizon is negative with:
\begin{equation}
|S_-| = 2\pi GM^2 \left(1- \sqrt{1-J^2/M^4G^2}   \right)\,.
\label{inner}
\end{equation}
 
Then the composition law in Eq.(\ref{BHTsallis}) gives the following total entropy of the Kerr black hole, which contains two separate contributions from the mass $M$ and from rotation with angular momentum $J$:
\begin{equation}
S_{\rm Kerr}(M,J)= 4\pi GM^2 + 4\pi J=S_{\rm Schwarzschild}(M) + 4\pi J
\,.
\label{KerrEntropy}
\end{equation}
Eq.(\ref{KerrEntropy}) suggests the quantization of the rotational part of the total entropy:
\begin{equation}
S_{\rm Kerr}(M,J)-S(M,0) =  4\pi \sqrt{J(J+1)}\,.
\label{quantization}
\end{equation}
In the thermodynamic limit $J\gg 1$ this gives
\begin{equation}
{\partial S}/{\partial J}{\big |}_M =  4\pi\,.
\label{dSdJ}
\end{equation}

If the Kerr black hole with $J\gg 1$ absorbs or emits a massless particle with spin $s_z=\pm 1/2$,
then the mass $M$ of the black hole remains the same, while its entropy changes by the following amount:
\begin{equation}
|\Delta S| = 2\pi\,.
\label{EntropyChange}
\end{equation}
Such stepwise behaviour of the black hole entropy is discussed in many papers, starting with Bekenstein who argued that entropy should be quantized in equidistant steps,\cite{Bekenstein1973,Bekenstein1974} see e.g. the recent paper\cite{Jana2025} and references therein.
But in Eq.(\ref{quantization}) only the rotational part of the entropy of RN black hole is quantized.

The entropy of the extremal Kerr black hole with $J=GM^2$ is
\begin{equation}
S_{\rm Kerr \, extreme}=S(M, J=GM^2) =2S(M, 0)= 8\pi J\,.
\label{KerrExtremal}
\end{equation} 
It is four times the traditionally discussed value $S_0=2\pi J$. 
Also, if Eq.(\ref{KerrExtremal}) is written in the form
\begin{equation}
S_{\rm Kerr\,extreme}= 8\pi \sqrt{J(J+1)}\,,
\label{KerrExtremal2}
\end{equation} 
  this corresponds to Eq.(11) in Ref. \cite{Khriplovich1998}  with an additional factor of 4. Instead of being two times smaller than the entropy of a Schwarzschild black hole of the same mass $M$, it is two times larger.

\section{de Sitter local thermodynamics}
\label{dSthermo}

The thermodynamics of de Sitter is essentially different from that of the black holes. It appears that it is not determined by the temperature of the Hawking radiation from the cosmological horizon.

\subsection{Hydrogen atom in de Sitter environment and de Sitter temperature}
\label{AtomInDS}

Let us show that matter perceives the de Sitter state of the quantum vacuum as the heat bath. For that  let us consider an atom  at the origin, $r = 0$. The atom plays the role of the detector (or the role of the static observer) in this spacetime. The electron bounded to an atom may absorb the energy from the gravitational field of the de Sitter background and escape from the electric potential barrier.  If the ionization potential is much smaller than the electron rest energy but is much larger than the Hubble parameter, $\hbar H\ll \epsilon_0 \ll mc^2$, one can use
the non-relativistic quantum mechanics to estimate the tunneling rate through the barrier. 

Let's consider an electron at the $n$-th level in a hydrogen atom. Under de Sitter gravitational field this electron can escape from the atom with the conservation of energy, which in the classical limit is given by the classical equation: 
\begin{eqnarray}
\frac{p_r^2}{2m} +p_rv(r) = -E_n \,\,,\,\, E_n= \frac{me^4}{2\hbar^2}\frac{1}{n^2}\,.
\label{Classical1}
\end{eqnarray}
 Here $v(r)=Hr$ and $p_r(r)v(r)$ is the Doppler shift, which allows for electron to reach the negative energy when it escapes from the atom.
 The corresponding radial trajectory $p_r(r)$ for escape of electron from the atom is:
\begin{eqnarray}
p_r(r)= -mv(r) + \sqrt{m^2v^2(r) -2m E_n}\,,
\label{ElectronTrajectory}
\end{eqnarray}
and this trajectory was used for calculation of ionization rate.

However, since the trajectory is well inside the horizon, one can can obtain the same result using the classical (i.e. non-relativistic) gravitational potential $U(r)=-mv^2(r)/2=-mH^2r^2/2$.\cite{Maxfield2022} 
Then the bound state decays by quantum tunnelling of electron from the point $r=0$ to the point $r=r_n$, at which the electron level  $-E_n$ matches the de Sitter gravitational potential,  $U(r_n)=-mH^2r_n^2/2=-E_n$. The radial trajectory $p_r(r)$ follows now from the classical equation 
\begin{equation}
\frac{{\bf p}^2}{2m}  - \frac{1}{2}mH^2r^2 = -E_n\,.
\label{GravPotential}
\end{equation}
One obtains
\begin{eqnarray}
p_r(r)= \sqrt{m^2H^2r^2 -2m E_n}\,,
\label{ElectronTrajectory2}
\end{eqnarray}
and the integral of $p_r(r)$ over the classically forbidden region, $0 < r < r_n=\sqrt{2E_n/mH^2}$, gives the  ionization rate 
\begin{eqnarray}
w\sim \exp{ \left(-\frac{2}{\hbar}\,{\rm Im}\, S\right)}=
\nonumber
\\
= \exp{ \left(-\frac{2}{\hbar}\int_0^{r_n} dr \sqrt{2m E_n-m^2H^2r^2}\right)} =
\label{WKB}
\\
=\exp\left(-\frac{\pi E_n}{\hbar H} \right) \equiv exp\left(-\frac{E_n}{T_{\rm dS}} \right) \,\,,\,\, T_{\rm dS}=\frac{\hbar H}{\pi}\,.
\label{IonizationRate}
\end{eqnarray}
The ionization rate is equivalent to the rate of ionization in the flat Minkowski spacetime in the presence of the heat bath with temperature $T_{\rm dS}=\hbar H/\pi$. 
This suggests that the de Sitter state of the quantum vacuum is the mixed state,\cite{Buchmuller2026} that serves as the heat bath for matter or for other degrees of freedom.

\subsection{Relation between de Sitter temperature and Gibbons-Hawking temperatures}
\label{LocalAndGH}

This heat bath temperature is twice the Gibbons-Hawking temperature  $T_{\rm GH}=H/2\pi$, which is generally considered as the temperature of the cosmological horizon describing the Hawking radiation from the horizon. Since the electron's trajectory is deep inside the horizon, $r_n \ll r_H=c/H$, the ionization process is fundamentally different from the process of Hawking radiation from the cosmological horizon. However, there is a relation between the ionization temperature $T_{\rm dS}$ and the Gibbons-Hawking temperature of Hawking radiation, $T_{\rm dS}=2T_{\rm GH}$. This follows from the symmetry of the de Sitter state.\cite{Volovik2024}
The reason for this is that the temperature $T_{\rm dS}$ determines not only local processes, but also the process of Hawking radiation from the cosmological horizon.

In the Hawking process in de Sitter state, two particles are created coherently: one particle is created inside the horizon, while its partner is simultaneously created outside the horizon. The rate of the coherent cotunneling of two particles, each with energy $E$, is determined by the temperature $T_{\rm dS}$, i.e. $w\propto \exp(-2E/T_{\rm dS})$. However, the observer who uses the Unruh-DeWitt detector can detect only the particle created inside the horizon. For this observer with limited information, the creation rate  $w\propto \exp(-2E/T_{\rm dS})$ is perceived as 
\begin{equation}
w\propto \exp\left(-\frac{E}{T_{\rm dS}/2} \right)=\exp\left(-\frac{E}{T_{\rm GH}} \right)\,.
\label{doubleT}
\end{equation}
That is why  $T_{\rm dS}=H/\pi$ is the relevant thermodynamic temperature, which determines both the local processes and the processes related to the cosmological horizon.

The similar situations with two physical processes should take place in the Unruh effect: the process related to Rindler horizon\cite{Unruh1976} and the local process. This requires the further consideration, see e.g. Ref. \cite{Volovik2024Unruh}.

\subsection{Local entropy of de Sitter, bulk entropy and entropy of Hubble volume}
\label{LocalGlobal}

The behaviour of matter in the de Sitter environment suggests that the de Sitter state can be considered as the heat bath, where the local temperature 
$T_{\rm dS}=H/\pi$ determines the thermodynamics of the de Sitter state.

From Friedmann equations of general relativity one can express the energy density of the de Sitter vacuum in terms of this local temperature, and then find its free energy and the entropy density.
The energy density, which is the cosmological constant $\Lambda$, is ($\hbar=c=1$):\cite{Volovik2024}
\begin{equation}
 \epsilon_{\rm vac}=\Lambda=\frac{3}{8\pi G}H^2=\frac{3\pi}{8G}T_{\rm dS}^2\,.
\label{dSEnergyDensity}
\end{equation}
This determines the free energy density $F$ of the de Sitter state. From equation $F- T_{\rm dS} dF/dT_{\rm dS}=\epsilon_{\rm vac}$ one obtains $F(T_{\rm dS})=-\epsilon_{\rm vac}(T_{\rm dS})$, and thus the entropy density $s_{\rm dS}$ is:
\begin{equation}
s_{\rm dS}= - \frac{\partial F}{\partial T_{\rm dS}} =\frac{3\pi}{4G}T_{\rm dS}=\frac{3}{4G}H\,.
\label{dSEntropyDensity}
\end{equation}
This is an example of the entropy density of spacetime suggested by Padmanabhan,\cite{Padmanabhan2010} see also review papers on the gravitational entropy.\cite{Guha2023,Ong2022}

The de Sitter state is homogeneous and thus the entropy of any part of the de Sitter state is proportional to the volume $V$ of this part:
  \begin{equation}
S(V)=s_{\rm dS}V   \,.
\label{dSvolume}
\end{equation}
The volume $V$ is arbitrary, it can be smaller or larger than the Hubble volume.

 The special case is presented by the entropy of the Hubble volume. It is obtained by
  integrating of the entropy density (\ref{dSEntropyDensity}) over the Hubble volume $V_{\rm Hubble}$:
  \begin{equation}
S_{\rm Hubble}=S(V_{\rm Hubble})=s_{\rm dS}V_{\rm Hubble}=\frac{A}{4G}  \,,
\label{dSHubble}
\end{equation}
where $A$ is the area of the cosmological horizon. The entropy in Eq.(\ref{dSHubble}) has the same form as the entropy of the black hole horizon, and it coincides with the entropy of the cosmological horizon  introduced by Gibbons and Hawking, $S_{\rm GH}=\frac{A}{4G}$.
This coincidence gives physical meaning to the Gibbons-Hawking entropy: it can be associated with the entropy of the region inside the cosmological horizon.  That is why we have the holographic bulk-horizon correspondence, in which the entropy of bulk is equal to the entropy of the surface. Note that such holography does not take place if one uses the Gibbons-Hawking temperature $T_{\rm GH}$ as the local temperature.

\subsection{Two-component de Sitter thermodynamics}
\label{TwoComponents}

Since the de Sitter state has local temperature, the local energy density $\epsilon_{\rm vac}$ and pressure $P_{\rm vac}=-\epsilon_{\rm vac}$, then we must have the corresponding thermodynamic laws.  The conventional expression for the Gibbs-Duhem relation, $T_{\rm dS}s_{\rm dS}=\epsilon_{\rm vac}+ P_{\rm vac}$, does not work, since $\epsilon_{\rm vac}+ P_{\rm vac}=0$.
The reason is that we did not take into account the gravitational degrees of freedom. 
Earlier it was shown, that gravity contributes to thermodynamics with the pair of the  thermodynamically conjugate variables:  the gravitational coupling $K=\frac{1}{16\pi G}$ and the Riemann curvature  ${\cal R}$, see Refs.\cite{KlinkhamerVolovik2008c,Volovik2022G,Volovik2020}. The emergence of the space independent thermodynamic variable ${\cal R}$ in de Sitter is an example of the so-called Kronecker anomaly,\cite{PolyakovPopov2022} or of the Larkin-Pikin effect,\cite{LarkinPikin1969} where the extra degrees of freedom emerge in the fully homogeneous state, i.e. at ${\bf k}=0$. The  Kronecker anomaly has been also applied to the Bonnor-Melvin-$\Lambda$ Universe and static Einstein Universe.\cite{Volovik2026Bonnor}

Since the scalar curvature in de Sitter is ${\cal R}=12H^2$, one obtains
 \begin{equation}
T_{\rm dS} s_{\rm dS}=  K{\cal R}\,.
\label{TSisKR}
\end{equation}
This demonstrates that it is the gravitational degrees of freedom that provide the entropy of the de Sitter state. 

So, we have two components of the de Sitter state, the dark energy component with 
equation of state $P_{\rm vac}=-\epsilon_{\rm vac}$, and the thermal gravitational component in 
Eq.(\ref{TSisKR}). 
This equation can be rewritten in the form of Gibbs-Duhem relation, which incorporates corresponding energy density $\epsilon_{\cal R}$ of the gravitational component and its pressure  $P_{\cal R}$.
 \begin{equation}
T_{\rm dS}s_{\rm dS}=  \epsilon_{\cal R}+ P_{\cal R}\,.
\label{GibbsDuhem}
\end{equation}
The gravitational pressure is obtained from the requirement, that in the absence of external pressure
the total pressure of the two components is zero, $P_{\cal R}=-P_{\rm vac}=\epsilon_{\rm vac}$. And then Eq.(\ref{GibbsDuhem}) gives 
 \begin{equation}
\epsilon_{\cal R} =\epsilon_{\rm vac}= P_{\cal R}\,,
\label{DarkDark}
\end{equation}
which means that the gravitational component of de Sitter has the equation of state $w_{\cal R}=1$.

This demonstrates that the gravitational component is similar to stiff matter introduced by Zel'dovich,\cite{Zeldovich1962} where the speed of sound is equal to the speed of light, $c_s^2=c^2 dP_{\cal R}/d\epsilon_{\cal R}=c^2$. 
In our case this mode represents the propagation of entropy, which exactly corresponds to the second sound in the Landau two-component model, where the dark energy plays the role of superfluid component, and gravity plays the role of the normal component, see Sec. \ref{SecondSound}. 

Finally we have for the two components of de Sitter:
 \begin{eqnarray}
P_{\rm vac}=w_{\rm vac}\epsilon_{\rm vac} \,\,,\,\, w_{\rm vac}=-1\,,
\label{wvac}
\\
P_{\cal R}=w_{\cal R}\epsilon_{\cal R}  \,\,,\,\, w_{\cal R}=1\,,
\label{wR}
\\
P_{\cal R}+P_{\rm vac}=0\,.
\label{pressures}
\end{eqnarray}

If the matter component is added, one would have the three-fluid hydrodynamics:
vacuum + gravity + matter.
The three-component thermodynamics of the static Einstein Universe see in Refs. \cite{Volovik2024e,Volovik2026Bonnor} and in Sec. 29.4.5 of Ref. \cite{Volovik2003}.

\subsection{de Sitter bubble}
\label{BubbleSec}

To explain the physical meaning of the external pressure acting on the de Sitter state, let us consider a de Sitter bubble of radius $R_{\rm bubble}$.\cite{KlinkhamerVolovik2017} Then external pressure is the pressure acting on this bubble, and we assume that it is zero. But now we must take into account the surface tension of the bubble boundary, $\sigma\sim E_{\rm P}^3$, where  $E_{\rm P}$ is the Planck energy scale, and the corresponding Laplace pressure $P_{\rm L} \sim E_{\rm P}^3/R_{\rm bubble}$. The Laplace pressure can be neglected compared to the vacuum pressure $P_{\rm vac}$ if the de Sitter bubble is large enough:
\begin{equation}
R_{\rm bubble} \gg \frac{E_{\rm P}}{H^2} \gg \frac{1}{H}\,.
\label{BubbleR}
\end{equation}
Only under this condition, when the radius of the de Sitter bubble highly exceeds the horizon radius $1/H$, will the partial pressures of gravitational and dark energy components compensate each other in Eq.(\ref{pressures}).

\subsection{Graviton in de Sitter as second sound in two-fluid thermodynamics}
\label{SecondSound}

Let us consider a particular consequence of the connection between cosmological two-fluid hydrodynamics and ordinary two-fluid hydrodynamics represented, for example, by superfluid $^4$He. We are interested in the analog of the second sound -- the propagating waves of the entropy density $S$, which is concentrated in the thermal normal components of the cosmological fluid. 

The velocity $s_2$ of the second sound in the two-fluid hydrodynamics has the following general form:\cite{Schmitt2014}
\begin{equation}
s_2^2=\frac{TS^2}{C_V\rho} \,\frac{\rho_s}{\rho_n} \,,
\label{EntropyRatio}
\end{equation}
where $C_V$ is the specific heat; and  $\rho=\rho_s +\rho_n$ is the liquid density.

In the de Sitter state one has $S=s_{\rm dS}$; $T=T_{\rm dS}$; $\epsilon_{\cal R} \equiv \rho_n c^2$; 
$\epsilon_{\rm vac} \equiv \rho_s c^2 =  \rho_n c^2$; $\rho =\rho_n +\rho_s=2\rho_n$; and $C_V=s_{\rm dS}$. Then from Eq.(\ref{EntropyRatio}) one obtains that the velocity of the second sound coincides with the speed of light:
\begin{equation}
 s_2=c \,.
\label{SpeedSecondSound}
\end{equation}

At first glance this seems strange: why would Landau's classical two-fluid hydrodynamics equations lead to a mode that propagates at the speed of light. But apparently the two-fluid approach is quite general. It follows from the general laws of thermodynamics, which apply also to the relativistic quantum vacuum. In particular, thermodynamics allows us to solve the cosmological constant problem.\cite{Volovik2025} The mechanism of cancellation of the vacuum energy in the ground state of the system is purely thermodynamic and does not depend on whether the vacuum is relativistic or not. The relativistic behaviour of the cosmological second sound supports the applicability of the two-fluid approach to the de Sitter state.

Since the thermal component of this two-fluid hydrodynamics is represented by the gravitational degrees of freedom (the scalar curvature ${\cal R}$), the second sound mode represents graviton propagating in de Sitter. It would be interesting to derive this massless de Sitter graviton directly from the de Sitter dynamics. However,  there is still no consensus regarding massless modes in the de Sitter background, see e.g. \cite{Garidi2003,Akhmedov2017,Gazeau2023,Sadekov2024,Hinterbichler2025,Glavan2025,Rajantie2025,Todorov2026}.

 \subsection{Two-fluid de Sitter thermodynamics, QCD vacuum and cosmological constant}
  \label{QCDSec}

The two-fluid de Sitter thermodynamics allows us to estimate the present value of the cosmological constant in the scenario, which is related to quark confinement. In this scenario, the dark energy is generated by the vacuum of Quantum Chromodynamics (QCD), see Ref. \cite{Zhitnitsky2026} and references therein. 
In the fully equilibrium ground state, i.e. in the Minkowski vacuum, all the contributions to the dark energy cancel each other. In the expanding de Sitter Universe, the cancellation is not complete, and in the QCD scenario the dark energy is proportional to the Hubble parameter $H$:\cite{Zhitnitsky2026} 
\begin{equation}
 \epsilon_{\rm vac}=c_H\Lambda_{\rm QCD}^3 H \,,
\label{QCDEnergyDensity}
\end{equation}
where $\Lambda_{\rm QCD} \sim 100$ MeV is the QCD energy scale and $c_H$ is the dimensionless parameter on the order of unity.
The same estimation was obtained\cite{KlinkhamerVolovik2009} using the Gribov picture of quark confinement.\cite{Gribov1078}   

In the two-fluid thermodynamics of de Sitter, the vacuum energy density $\epsilon_{\rm vac}$ coincides with the energy density $\epsilon_{\cal R}$ of the gravitational degrees of freedom, see Eq.(\ref{DarkDark}). Then from equation
$\epsilon_{\rm vac}= \epsilon_{\cal R}=6KH^2$ and Eq.(\ref{QCDEnergyDensity}), one obtains the following estimation of the present value of the Hubble parameter and correspondingly the present value of the cosmological "constant":  
\begin{equation}
H=c_H\frac{\Lambda_{\rm QCD}^3}{6K}  \,\,,\,\,
\Lambda\equiv \epsilon_{\rm vac}  = c_H^2 \frac{\Lambda_{\rm QCD}^6}{6K} \,.
\label{QCDEnergyDensity2}
\end{equation}
Since the QCD scale $\Lambda_{\rm QCD}$ is responsible for the proton mass $m_p$, this estimation of the cosmological constant corresponds to Zeldovich’s original suggestion,\cite{Zeldovich1967,Zeldovich1968}  $\Lambda \sim Gm_p^6$. 

The value of $\Lambda$ in the equation (\ref{QCDEnergyDensity2}) may represent one of several possible dark energy plateaus in the expanding Universe,\cite{KlinkhamerVolovik2011} but finally 
 $\Lambda$ relaxes to zero.

  \section{Entropy of cosmological horizon in de Sitter (d+1)-spacetime}
  \label{ModEntropySec}

 \subsection{On-shell vs off-shell}
  \label{OnshellSec}
  
  To emphasize the connection between the entropy of the Hubble volume (obtained by integration over the volume inside the cosmological horizon)  and entropy related to the area $A$ of horizon,
  Diakonov\cite{Diakonov2025} introduced the notions of the on-shell and off-shell entropies. The first represents the entropy of the Hubble volume, and the second corresponds to the entropy traditionally associated with the cosmological horizon.
      
 Let us consider this connection assuming that the local temperature of de Sitter is $T=aH$. Here we do not specify the parameter $a$. The local processes, such as ionization of an atom in de Sitter environment,\cite{Volovik2025F,Maxfield2022}  suggest $a=1/\pi$, while the Hawking radiation from the cosmological horizon suggests $a=1/2\pi$.  
  
Let us first consider the on-shell entropy. The temperature dependence of energy density in the de Sitter $d+1$ dimensional spacetime is given by (here we use $G=1$):
  \begin{equation}
\epsilon=\frac{d(d-1)}{16\pi} H^2 =\frac{d(d-1)}{16\pi} \frac{T^2}{a^2}
\,.
\label{EnergyDensity}
\end{equation}
This energy density determines the local entropy density, $s=-\frac{dF}{dT}$, where $F$ is the free energy density. For the $T^2$ dependence of the energy density, one has $F=-\epsilon$, and thus the entropy density is
 \begin{equation}
s =\frac{d\epsilon}{dT}=\frac{d(d-1)}{8\pi} \frac{T}{a^2} = \frac{d(d-1)}{8\pi} \frac{H}{a}
\,,
\label{EntropyDensity}
\end{equation}

The de Sitter state is homogeneous, and its entropy is extensive, i.e. it is proportional to the considered volume, $S_V=sV$. Here the volume $V$ is arbitrary, it can be smaller and larger than
the Hubble volume $V_H$, which is:
 \begin{equation}
V_H=\frac{\pi^{d/2}}{\Gamma(\frac{d+2}{2})} \frac{1}{H^d}
\,.
\label{HubbleVolume}
\end{equation}
The on-shell entropy is the entropy of the Hubble volume:
 \begin{equation}
 s V_H= \frac{d(d-1)}{8\pi} \frac{\pi^{d/2}}{\Gamma(\frac{d+2}{2})} \frac{1}{H^{d-1}a}
 = \frac{d-1}{4\pi} \frac{\pi^{d/2}}{\Gamma(\frac{d}{2})} \frac{1}{H^{d-1}a}
\,.
\label{HubbleEntropy}
\end{equation}

Let us compare this on-shell entropy of the Hubble volume with the off-shell entropy, which is traditionally associated with the area law for the Gibbons-Hawking entropy of the cosmological horizon:
\begin{equation}
S_{\rm GH}=\frac{A}{4}=  \frac{\pi^{d/2}}{2\Gamma(\frac{d}{2})} \frac{1}{H^{d-1}}
\,,
\label{HorizonEntropy}
\end{equation}

The on-shell and off-shell entropies coincide, $sV_H=S_{\rm GH}$, if $a=\frac {2}{(d-1)\pi}$, i.e. if the local temperature depends on dimension:
\begin{equation}
T=\frac {2}{(d-1)\pi} H
\,.
\label{LocalT}
\end{equation}
For $d=3$ this temperature agrees with the local temperature, which is obtained by consideration of the rate of all the local activation processes in the de Sitter environment,  $T(d=3)=T_{\rm dS}=H/\pi$. 
But for the other dimensions, $d\neq 3$, the Eq.(\ref{LocalT}) does not agree with the local temperature. However, we already demonstrated that the temperature $T=T_{\rm dS}=H/\pi$ is the proper thermodynamics temperature, which determines both the local processes and processes related to the cosmological horizon. This temperature does not depend on dimension $d$ and thus we must keep $T_{\rm dS}=H/\pi$.  But if we keep both the local temperature $T=T_{\rm dS}=H/\pi$  and the standard Gibbons-Hawking entropy $A/4$ in Eq.(\ref{HorizonEntropy}),  then the holographic bulk-horizon correspondence is violated.

That is why, the only way to reconstruct the holography, i.e. to relate the off-shell entropy of the horizon to the on-shell entropy of the region bounded by the horizon and preserve the local temperature $T(d)=T_{\rm dS}=H/\pi$, the Gibbons-Hawking entropy must be modified: 
\begin{equation}
S_{\rm GH}(d)= \frac{d-1}{8}A  \,\,,\,\, T(d)=T_{\rm dS} = \frac{H}{\pi}
\,.
\label{GHmodified}
\end{equation}

In principle, this modification of  the Gibbons-Hawking entropy  is not very unusual. Even the entropy of a black hole can deviate from $A/4$, examples of which are the entropy of the Reissner-Nordstr\"om black hole\cite{Volovik2025TC}  and the entropy of Kerr black hole in Eq.(\ref{BHTsallis}). The RN and Kerr black holes have two horizons, and thus the traditional approach, in which the entropy is determined by the area of the outer horizons, is not applicable.

In the next section we confirm the modified Gibbons-Hawking entropy (\ref{GHmodified})
of the cosmological horizon using the first law of thermodynamics. Thus the modified Gibbons-Hawking entropy for $d\neq 3$ is another example where the entropy of the horizon is not equal to $A/4$, although for the cosmological horizon it remains proportional to the area.

\subsection{First law of thermodynamics of de Sitter horizon}
\label{FirstLawSec}

In Sec. \ref{TwoComponents}, we considered the local thermodynamics of de Sitter with the Gibbs-Duhem relation in Eq.(\ref{GibbsDuhem}).  Now let us consider the thermodynamics of the Hubble volume $V_H$ -- the part of the de Sitter space inside the cosmological horizon. Since $V_H$ depends on temperature, the first law of thermodynamics must include the pressure term. For the thermal gravitational component of the de Sitter thermodynamics, with $T=T_{\rm dS}$, $P=P_{\cal R}$, $E_H=\epsilon_{\cal R} V_H$, and $S_H=s_{\rm dS}V_H$, the first law of thermodynamics has the conventional form:
  \begin{equation}
TdS_H = dE_H + P  dV_H \,.
\label{FirstLaw1}
\end{equation}

Let us check this equation considering first the dimension $d=3$. The quantities which enter the  first law in Eq.(\ref{FirstLaw1}) for $S_H=A/4$ are:
  \begin{eqnarray}
TdS_H =T\frac{dA}{4}= T d \left(\frac{\pi}{H^2} \right)=-2\pi T\frac{dH}{H^3}\,,
\label{dSH}
\\
dE_H= d\left(\frac{4\pi}{3H^3}\frac{3H^2}{8\pi} \right) = d\left(\frac{1}{2H}\right)
=-\frac{1}{2}\frac{dH}{H^2}\,,
\label{dEH}
\\
P  dV_H= \frac{3H^2}{8\pi} d\left(\frac{4\pi}{3H^3}\right) = -\frac{3}{2}\frac{dH}{H^2} \,.
\label{dVH}
\end{eqnarray}
Inserting these terms to Eq.(\ref{FirstLaw1}), one obtains that  the first law is satisfied if $T$ is the Sitter  local temperature, $T=T_{\rm dS}=H/\pi$. This supports the Gibbons-Hawking entropy $S_{\rm GH}=A/4$ for $d=3$. The same result is valid in the $f({\cal R})$-theory.\cite{Volovik2024}

 \subsection{Cosmological horizon entropy in other dimensions}
 \label{dSother}
 
Let us now find the entropy of the horizon at arbitrary dimension $d$. As follows from the probability of the activation processes in de Sitter environment, the local temperature $T=H/\pi$ is universal: it does not depend on $d$ and is always twice the Gibbons-Hawking temperature. So, we have:
  \begin{eqnarray}
TdS_H =\frac{H}{\pi}dS_H\,,
\label{dSHd}
 \end{eqnarray}
 while
 \begin{eqnarray}
dE_H +P  dV_H=-\frac{(d-1)^2}{4\pi}  \frac{\pi^{d/2}}{\Gamma(\frac{d}{2})} \frac{dH}{H^{d-1}} \,.
\label{dEHPdV}
 \end{eqnarray}
 
Equating equations (\ref{dSHd}) and (\ref{dEHPdV}), we obtain
 \begin{eqnarray}
dS_H=-\frac{(d-1)^2}{4}  \frac{\pi^{d/2}}{\Gamma(\frac{d}{2})} \frac{dH}{H^{d}} \,,
\label{dEHPdV}
 \end{eqnarray}
 which gives for the entropy of the Hubble volume in space dimension $d$:
  \begin{eqnarray}
S_H(d)=\frac{d-1}{4}  \frac{\pi^{d/2}}{\Gamma(\frac{d}{2})}  \frac{1}{H^{d-1}} = \frac{d-1}{2} \frac{A}{4}\,.
\label{dEHPdV}
\end{eqnarray}
Eq.(\ref{dEHPdV}) coincides with Eq.(\ref{GHmodified}) for the modified Gibbons-Hawking entropy. 
Thus, the first law of de Sitter thermodynamics with the local temperature $T=T_{\rm dS}=H/\pi$ gives a modification of the Gibbons-Hawking entropy of the cosmological horizon in Eq.(\ref{GHmodified}). Note that this conclusion does not require accepting the controversial sign-reversal proposals;\cite{Jacobson2023,Diakonov2025}
it follows solely from local de Sitter thermodynamics, which includes the gravitational pair of the thermodynamically conjugate variables: the gravitational coupling $K$ and scalar curvature ${\cal R}$.

 \subsection{Negative entropy and contracting de Sitter}
 \label{contracting}
 
The controversial negative sign in the off-shell approach in Ref.\cite{Diakonov2025} and in the first law for the de Sitter horizon in Ref.\cite{Jacobson2023} actually corresponds to a negative sign for the entropy, and this violates the bulk-horizon correspondence. In the local thermodynamic approach the contradiction does not arise: the signs of both entropies (of the Hubble volume and of cosmological horizon) coincide. 

However, the negative entropy may also take place, but this happens for the contracting de Sitter state. Both the entropy of the Hubble volume and the entropy of cosmological horizon are negative for the contracting Universe. This becomes clear, when the Painlev\'e-Gullstrand (PG) coordinates \cite{Painleve,Gullstrand}  are used, which have no singularity at the horizon.
The PG metric is given by
 \begin{eqnarray}
ds^2= - c^2dt^2 +   (d{\bf r} - {\bf v}({\bf r})dt)^2 \,,
\label{G1}
\end{eqnarray}
where ${\bf v}({\bf r})=H{\bf r}$ is the shift velocity. The positive Hubble parameter, $H>0$, corresponds to expansion and the negative Hubble parameter, $H<0$, corresponds to contraction. 
Similar situation for the entropies of black and white holes was discussed in Section \ref{WhiteHoleSec}.

For the contracting de Sitter Universe, the local temperature is negative, $T=H/\pi <0$. This produces the negative entropy density in Eq. (\ref{EntropyDensity}):
 \begin{equation}
s =\frac{d\epsilon}{dT}=\frac{\pi d(d-1)}{8} T= \frac{d(d-1)}{8} H <0
\,,
\label{EntropyDensityNeg}
\end{equation}
and correspondingly the negative entropy of the Hubble volume:
 \begin{eqnarray}
S_H=\frac{d-1}{4}  \frac{\pi^{d/2}}{\Gamma(\frac{d}{2})}  \frac{H}{|H|^d} <0\,.
\label{NegEntropy}
\end{eqnarray}

This agrees with the negative sign of the entropy of the cosmological horizon in the contracting de Sitter. This cosmological horizon is the white horizon, in the same way as the horizon of the Schwarzschild white hole and the inner horizon of the RN black hole.\cite{Volovik2025TC} The black and white horizons are distinguished using the PG metric instead of  static coordinates. Consider the coordinate dependence of the shift velocity $v(x)$ across the horizon at $x=0$.
For the black horizon one has $dv/dx|_{x=0} >0$, while for the white horizon $dv/dx|_{x=0} <0$.
Applying this to the de Sitter horizon, one obtains $dv/dr |_{r=r_H} =H>0$ for the expanding de Sitter 
and $dv/dr |_{r=r_H} =H<0$ for the contracting de Sitter.

The white horizons have negative entropies, as follows from the consideration of the Hawking radiation. This demonstrates that for the contracting de Sitter, the entropy of horizon is again equal to the entropy $S_H$ of the region inside the cosmological horizon. But now both entropies are negative. The first law (\ref{FirstLaw1}) of de Sitter thermodynamics is also valid for the contracting de Sitter, since both entropy and temperature change sign.

 \subsection{Possible connection to Wald entropy}
 \label{WaldSec}

Let us consider whether it is possible to obtain the modified Gibbons-Hawking entropy through the Wald scenario of dynamic entropy:\cite{Wald1993} 
 \begin{equation}
S_{\rm Wald}(d)=\left(1- \alpha \partial_\alpha  \right)S(\alpha)|_{\alpha=1}
\,.
\label{Wald1}
\end{equation}
The goal is to obtain the modifed Wald entropy in the form:
 \begin{equation}
S_{\rm Wald}(d)= \frac{d-1}{8} A
\,.
\label{Wald3}
\end{equation}
For that one must assume that
 \begin{equation}
S(\alpha)=\alpha^{(3-d)/2} \frac{A}{4} \,.
\label{Wald2}
\end{equation}
The modified Wald entropy is still proportional to the area $A$, but the question of justifying the equation (\ref{Wald2}) remains open.

 \subsection{Thermodynamic and quantum fluctuations of horizon area}
 \label{fluctuations}

The connection between macroscopic quantum tunnelling and entropy in systems with event horizon in Eq.(\ref{HoleEmession2}) also suggests the connection between quantum and thermodynamic fluctuations in these systems. The variance of the thermal entropy corresponds to the quantum fluctuations of the area of the horizon.\cite{Volovik2025fluct}

The de Sitter state is homogeneous and isotropic, so we can apply the standard equations for the thermodynamic fluctuations.\cite{Landau_Lifshitz} The local entropy density of the de Sitter state is linear in temperature, $s(T) \propto T$, and thus the heat capacity is also linear in $T$ . As a result one obtains the following fluctuations of the local entropy:
 \begin{equation}
\left<( \Delta s)^2\right> = \frac{s}{V} \,.
\label{LocalEntropyF}
\end{equation}
Then for entropy $S(V)=sV$ of the arbitrary volume $V$ in Eq.(\ref{dSvolume}) one has:
 \begin{equation}
\left<( \Delta S)^2\right> = \left<S\right> \,.
\label{EntropyF}
\end{equation}
The Eq.(\ref{EntropyF}) does not contain the Planck constant $\hbar$, which demonstrates that it describes thermodynamic fluctuations.  This equation also does not contain the gravitational coupling $K$. This is in agreement with the observation by Jacobson \cite{Jacobson1994} that both $K$ and $S$ are renormalized by the very same quantum fluctuations. 
Equation (\ref{EntropyF}) is automatically valid also for the entropy of the Hubble volume with $V=V_H$. A similar result with a factor of two difference is obtained from fluctuations of the Euclidean volume.\cite{Blanco2026} The difference may be related to a factor of two difference in the de Sitter temperature.

Let us first consider space dimension $d=3$, where $S(V_H)=sV_H=A/4G\hbar=4\pi KA/\hbar$. For this dimension one obtains the following fluctuations of the area of the cosmological horizon:
\begin{equation}
\left<( \Delta A)^2\right> = 4G\hbar \left<A\right>=\frac{\hbar \left<A\right>}{4\pi K} \,.
\label{AreaFCosm}
\end{equation}

 This equation for the area fluctuations contains the Planck constant $\hbar$. This suggests that thermal fluctuations of the horizon can be also considered as quantum fluctuations. To see that and find the corresponding canonically conjugate variables, let us take into account that the gravitational coupling $K$ is the special thermodynamic variable which comes from the Kronecker anomaly.\cite{PolyakovPopov2022} The variations of $K$ and $A$ at fixed energy of the Hubble volume obey the following relation:
\begin{equation}
\frac{\Delta A}{A}= 2\frac{\Delta K}{K} \,.
\label{ratio}
\end{equation}
Inserting this to Eq.(\ref{AreaFCosm}) one obtains a kind of the quantum uncertainty relation:
\begin{equation}
\Delta A \,\Delta K= \frac{\hbar}{8\pi} \,.
\label{uncertainty}
\end{equation}
This suggests that $\tilde K = 4\pi K$ is the variable, which is canonically conjugate to the area $A$ of the cosmological horizon. The same works for the black hole horizon.\cite{Volovik2025fluct}
These canonically conjugate variables, $\tilde K$ and the black hole area $A$,  were used for the calculations of the transition rate from the black hole to the white hole with the same mass.\cite{Volovik2022} This rate is determined by the integral $\int_C A(\tilde K)d\tilde K$ over the tunneling trajectory $C$, which connects black and white holes.

For general dimension $d\geq 3$, where $S(V_H)=2\pi (d-1)KA/\hbar$, the Eq.(\ref{AreaFCosm}) becomes
\begin{equation}
\left<( \Delta A)^2\right> = \frac{8}{d-1}G\hbar \left<A\right> =\frac{\hbar }{2\pi (d-1)K} \left<A\right> \,,
\label{AreaFCosmD}
\end{equation}
Note that the coefficient in the area variance differs from the corresponding coefficients  in Refs. \cite{Ciambelli2025,ParikhPereira2024}.
Then the Eq.(\ref{ratio}) becomes
\begin{equation}
\frac{\Delta A}{A}= \frac{d-1}{d-2}\frac{\Delta K}{K} \,.
\label{ratioD}
\end{equation}
and one obtains the quantum uncertainty relation for general dimension $d$:
\begin{equation}
\Delta A \,\Delta K= \frac{\hbar}{2\pi} \frac {d-2}{(d-1)^2}\,.
\label{uncertaintyD}
\end{equation}

This means that $\tilde K= 2\pi K(d-1)^2/(d-2)$ is a variable canonically conjugate to the area $A$ of horizon. On the other hand, the gravitational coupling $K$ and the de Sitter curvature ${\cal R}$  are conjugate thermodynamic variables. The reason of the connection between quantum and thermal variables and correspondingly between the quantum and thermal fluctuations is the presence of the event horizon, which is a source of irreversibility, analogous to statistical irreversibility.\cite{Witten2025}

 \subsection{Combining black hole horizon and de Sitter horizon}
 \label{CombineSec}
 
The entropy of the homogeneous de Sitter state is extensive, $S(V_1+V_2)=S(V_1) +S(V_2)$, while the entropy of the compact black hole obeys the non-extensive Tsallis-Cirto composition rule in Eq.(\ref{TCentropy2}). However, there is the connections between these systems, if only the horizon entropy is considered. Example is provided by the black hole in the de Sitter environment -- the Schwarzschild-de Sitter (SdS) black hole, which has both the black hole horizon and the cosmological horizon. Let us use the approach based on the singular coordinate transformations, which in particular was used  for the calculations of the probability of the macroscopic quantum tunneling from black hole to white hole.\cite{Volovik2022} Since the coordinate transformation has a singularity, invariance under diffeomorphisms does not necessarily apply, and hence the coordinate transformation connects two  physically different objects -- the black hole with mass $M$ and the white hole with the same mass.
 
The SdS black hole with the shift vector $v^2(r)=H^2r^2 +\frac{2MG}{r}$ has two horizons: the black hole horizon at $r=r_-$ and the cosmological horizon at $r=r_+>r_-$. Let us  make the singular coordinate transformation $r\rightarrow -r$. Then one obtains the state with  $v^2(r)=H^2r^2 - \frac{2MG}{r}$, which corresponds to the negative mass of black hole. This state has only a single horizon at $r_0=r_+ + r_-$. The entropies of the initial and final states obey the composition law:
\begin{equation}
\sqrt{S(r_0)}= \sqrt{S(r_-)} +\sqrt{S(r_+)}\,.
\label{SdScomposition}
\end{equation}
This corresponds to the Tsallis-Cirto $\delta=2$ statistics and justifies the dS Matrix Theory suggested by Susskind.\cite{Susskind2023} One might expect that in the general dimension $d$, the  the Tsallis-Cirto parameter is  $\delta=\frac{d-1}{d-2}$, and the entropy of cosmological horizon becomes extensive in the limit $d\rightarrow \infty$.

So, the statistics related to the  black hole horizon and to the cosmological horizons coincide.
What about the holographic bulk-horizon correspondence that exists for the entropy of cosmological horizon? Does it apply to the black hole horizon? Unruh's answer is no, a black hole is not a container of entropy.\cite{Unruh2026} However, there are arguments based on the first law of thermodynamics that all the entropy of an ordinary Schwarzschild black hole is concentrated in the black hole singularity, while  the holographic correspondence is satisfied.\cite{Volovik2025F}
Note that Unruh's objections are applicable to the Srinivasan and Padmanabhan tunneling approach to the Hawking radiation.\cite{Padmanabhan1999} The reason is that Srinivasan and Padmanabhan used the static Schwarzschild coordinates, that are singular at the horizon and the inside and outside are causally disconnected regions. Moreover, the static Schwarzschild coordinates describe a static hole, which is an intermediate state between a black hole with entropy $S_{\rm BH}=A/4G$ and a white hole with the opposite shift vector and negative entropy $S_{\rm WH}=-A/4G$.\cite{Volovik2022G} The static hole  has zero entropy, $S_{\rm static}=0$.

In principle, the horizonless regular black holes with zero entropy are not excluded.\cite{Dymnikova1992,MazurMottola2023,Fauzi2025,Kamenshchik2026}

  \section{Conclusion}

In the de Sitter state in the $3+1$ dimensional spacetime, there is the holographic connection between the entropy of the Hubble volume and the Gibbons-Hawking entropy $S_{\rm GH}=A/4G$ of cosmological horizon. This gives physical meaning and a natural explanation to the Gibbons-Hawking entropy  -- it is the entropy in the volume $V_H$ bounded by the cosmological horizon.
If so, the same holographic bulk-horizon correspondence is to be valid for an arbitrary spacetime dimension $d+1$. 

Calculating the entropy $S_H$ of the Hubble volume $V_H$ in the $d+1$ de Sitter state, we obtained $S_H(d)=\frac{d-1}{8}A/G$. This suggests the modification of the Gibbons-Hawking entropy of the de Sitter cosmological horizon. It must be the same as the entropy of the Hubble volume, $S_{\rm GH}(d)\equiv S_H(d)=\frac{d-1}{8}A/G$.

This derivation was also confirmed using the first law of the de Sitter thermodynamics. In both cases we used the local thermodynamics of the de Sitter state with the local temperature $T=H/\pi$, which determines the rate of activation processes in the de Sitter environment.

The same approach to the contracting de Sitter state, i.e. with the negative Hubble parameter $H<0$, demonstrates that the corresponding Gibbons-Hawking entropy of the cosmological horizon is with minus sign the modified Gibbons-Hawking entropy of the horizon of the expanding Universe, $S_H(d)=-\frac{d-1}{8}A/G$. The minus sign of the entropy follows both from the local thermodynamics of the contracting de Sitter and from the properties of the white horizon in the contracting de Sitter.

We also demonstrated that the local thermodynamics of de Sitter  looks similar is the Landau two-fluid hydrodynamics, where the role of the superfluid component is played by the dark energy, while gravitational degrees of freedom (the scalar curvature ${\cal R}$) are responsible for the thermal component of the de Sitter cosmological liquid. In particular, it is shown that the graviton propagating in the de Sitter spacetime corresponds to the second sound -- the propagating mode of the entropy density. The graviton velocity is obtained directly from the classical Landau equation for the velocity of the second sound, which demonstrates the universality of the Landau two fluid hydrodynamics.

I thank Dmitrii Dyakonov for correspondence.

\end{document}